\documentclass[%
 reprint,
 amsmath,amssymb,
 aps,
prb,
floatfix,
]{revtex4-2}

\usepackage{graphicx}% Include figure files
\usepackage{dcolumn}% Align table columns on decimal point
\usepackage{bm}% bold math
\usepackage{enumitem}
\usepackage[normalem]{ulem}
\usepackage{xcolor}
\newcommand{\ga}{\gamma}
\newcommand{\PT}{\mathcal{PT}}

\begin{document}

\preprint{APS/123-QED}

\title{Edge-controlled non-Hermitian skin effect in the modified Haldane model}% Force line breaks with \\

\author{Nobuhiro Ito}
\affiliation{
Department of Electronic and Physical Systems,
Waseda University, Tokyo 169-8555, Japan
}
% \altaffiliation[Also at ]{Department of Electronic and Physical Systems, Waseda University, Tokyo 169-8555, Japan}%Lines break automatically or can be forced with \\
 \email{itonobuhiro@fuji.waseda.jp}
 
\author{Shun Uchino}%
\affiliation{
Department of Electronic and Physical Systems,
Waseda University, Tokyo 169-8555, Japan
}
\affiliation{%
    Department of Materials Science, Waseda University, Tokyo 169-8555, Japan
}%
 \email{shun.uchino@waseda.jp}

\date{\today}% It is always \today, today,

\begin{abstract}
The hybrid skin–topological effect (HSTE) arises from the interplay between the non-Hermitian skin effect and topologically protected edge states. 
Here, we investigate the HSTE associated with antichiral edge states. As a minimal model, we study a modified Haldane nanoribbon in which gain and loss are applied exclusively at the zigzag edges.
We show that in antichiral systems, the HSTE originates from the interplay between edge modes and counterpropagating bulk modes, a mechanism distinct
from that in conventional chiral systems.
Remarkably, in sufficiently narrow ribbons, 
applying gain or loss to only one edge can induce a skin effect
in states localized at the opposite edge. 
We refer to this phenomenon as
a nonlocal antichiral skin effect.
We further show that, in narrow systems, edge-localized dissipation can induce 
bulk skin modes when parity-time ($\PT$) symmetry is broken,
while the bulk non-Hermitian skin effect is forbidden
in the $\PT$-symmetric regime.
These results imply that edge-localized gain and loss can control both the emergence and the localization direction of the bulk skin modes.
 Our findings establish a symmetry-based mechanism for controlling non-Hermitian
 skin effects via edge dissipation in antichiral systems.
\end{abstract}

\maketitle

%\tableofcontents

\section{\label{sec:introduction}Introduction}
In recent years, non-Hermitian systems have attracted considerable attention 
owing to their unconventional spectral and topological properties~\cite{nonhermitianreview1,nonhermitianreview2,nonhermitiantopology,molignini2}. A hallmark of non-Hermitian physics
is the non-Hermitian skin effect (NHSE), in which eigenstates extended throughout the bulk under periodic boundary conditions (PBC) become localized at the boundary 
under open boundary conditions (OBC)~\cite{winding3,winding4,InnerSkin,yang1,yang2,yang3,yang4,balazs,molignini1}. The NHSE is guaranteed
by nontrivial point-gap topology characterized by spectral winding
in the complex-energy plane, reflecting a fundamental breakdown of conventional
bulk-boundary correspondence~\cite{wang1,wang2}.
\par
In systems with line-gap topology, non-Hermiticity can give rise to the hybrid skin–topological effect (HSTE)~\cite{HSTEreview,HSTEreview2}. In the HSTE, topologically protected edge states acquire additional localization due to the NHSE.
So far, HSTEs have been extensively studied in systems hosting
chiral or helical edge states, including models with 
asymmetric hopping~\cite{nonreHSTE1,nonreHSTE2,nonreHSTE3}, bulk gain and loss~\cite{2022PRB_HSTE,2022PRL_HSTE,largech,hyperbolic,anderson,antichiralHSTE,AnomalousHSTE}, and edge-localized dissipation~\cite{Ma2024PRR,edgegl1,edgegl2}. 
The experimental realizations have been reported in photonic crystals~\cite{PhC1,PhC2}, electrical circuits~\cite{circuit1,circuit2}, phononic crystals~\cite{phononic1,phononic2}, and active matter systems~\cite{activematter}. 
\par
\begin{figure}[tb]
    \centering
    \includegraphics[width=0.99\linewidth]{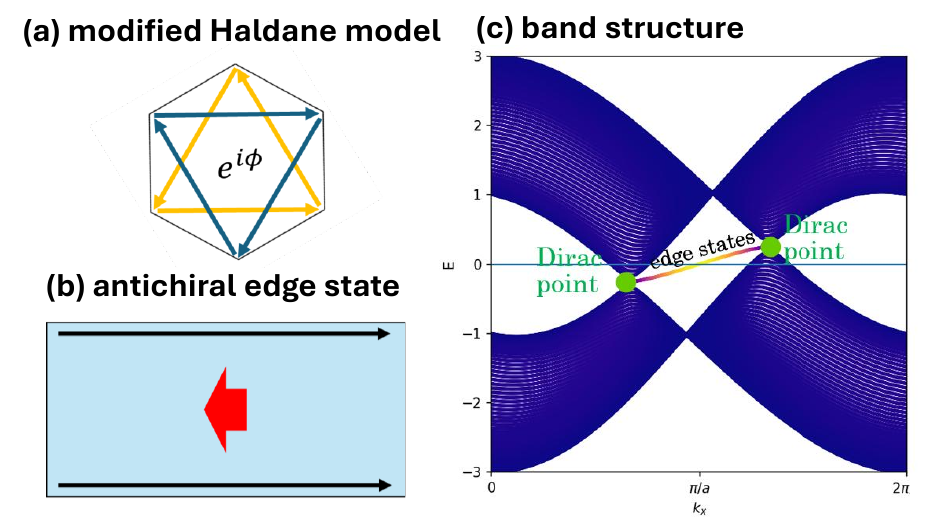}
    \caption{(a) Direction of phase accumulation in the NNN hopping of the modified Haldane model. (b) 
    Schematic illustration of antichiral edge states, which propagate in the same direction along opposite edges while counterpropagating modes reside in the bulk. (c) Band structure of the modified Haldane nanoribbon.}
    \label{fig:anti}
\end{figure}
Recently, topologically protected antichiral edge states were discovered in the \textit{modified Haldane model}~\cite{MHM}, which is obtained from
the celebrated Haldane model~\cite{Haldane} by
reversing the phases $\phi$ of the next-nearest-neighbor (NNN) hoppings  on one sublattice~[Fig.~\ref{fig:anti}(a)]. 
This phase pattern shifts the two Dirac points of the honeycomb lattice, resulting in the band structure shown in Fig.~\ref{fig:anti}(c). The zigzag edges of the honeycomb lattice intrinsically host edge states. 
The displacement of the Dirac points makes
 these edge states dispersive and gives them
 a finite group velocity, thereby 
turning them into antichiral edge states~\cite{nhanti1,nhanti2,anti1,anti2,anti3,anti4,anti5,anti6,anti7,anti8,anti9,anti10}.
The modified Haldane model has been realized in photonic crystals~\cite{antiPhC} and electrical circuits~\cite{anticircuit}.
Unlike conventional chiral edge states, antichiral edge states propagate in the same 
direction along the two opposite edges. To compensate for the resulting net current, counterpropagating modes appear in the bulk [Fig.~\ref{fig:anti}(b)]. 
\par
These features make antichiral systems
qualitatively different from conventional
chiral systems in the context of non-Hermitian topology. In particular, antichiral edge states are necessarily accompanied by counterpropagating bulk modes, suggesting that any HSTE in such systems should involve not only edge-state localization but also their coupling to bulk transport. Despite this distinctive structure,
HSTE in antichiral systems remains largely unexplored.
Although an HSTE associated with antichiral edge states has been predicted in an exciton-polariton system~\cite{antichiralHSTE}, its detailed mechanism remains poorly understood.

Motivated by this gap, we identify  the mechanism underlying the HSTE of antichiral edge states. The HSTE of antichiral edge states arises from the interplay between the edge states and these counterpropagating bulk states. This mechanism is distinct from those of HSTEs associated with other types of edge states reported in previous studies.
Remarkably, in sufficiently narrow ribbons, we find that introducing gain or loss exclusively to one edge induces localization of antichiral skin modes at the opposite edge. 
This phenomenon originates from the unique nature of antichiral systems,
and we refer to it as the \textit{nonlocal antichiral skin effect}.
\par
As a concrete platform with antichiral edge states, we focus on the modified Haldane model. To investigate its non-Hermitian properties, we 
consider a narrow  nanoribbon geometry and
introduce gain and loss exclusively at the zigzag edges. 
\par
Because the nanoribbon considered in this work is only a few atoms wide, the usual distinction between “bulk” and “edge” states becomes rather blurred. In such a narrow system, there is effectively no interior region that is free from the influence of the edges. 
Accordingly, edge-localized gain and loss induce the NHSE not only in the edge states but also in a large number of bulk states.
We further show that bulk non-Hermitian skin modes are forbidden in the
$\PT$-symmetric regime and emerge only when $\PT$ symmetry is broken. 
This result is consistent with a recent study
showing that $\PT$ symmetry breaking can activate 
the NHSE~\cite{PTNHSE}.
We also point out that edge-applied gain and loss can systematically control both the emergence and the localization direction of the bulk skin modes.
\par
This paper is organized as follows.
In Sec.~\ref{sec:model}, we introduce the modified Haldane nanoribbon with edge-localized gain and loss, and define the model studied in this work. In Sec.~\ref{sec:HSTEanti}, we introduce the mechanism of HSTE in antichiral systems. In Sec.~\ref{sec:result}, we present the numerical results and analyze the non-Hermitian skin effect in both antichiral edge states and bulk states, clarifying the role of $\PT$ symmetry. In Sec.~\ref{conclusion}, we summarize our findings and discuss their implications.
Detailed analyses of the $\PT$-broken regime, 
supplementary interpretations of the numerical results,
the bulk NHSE in additional bands, and numerical results for a sufficiently large system are presented in Appendixes~\ref{appenA}, \ref{appen:NHSEinter}, \ref{appenB}, and \ref{appenlargeNy}.

\section{\label{sec:model}model}
\subsection{\label{subsec:hamiltonian} Effective Hamiltonian}
\begin{figure}[htb]
    \centering
    \includegraphics[width=0.99\linewidth]{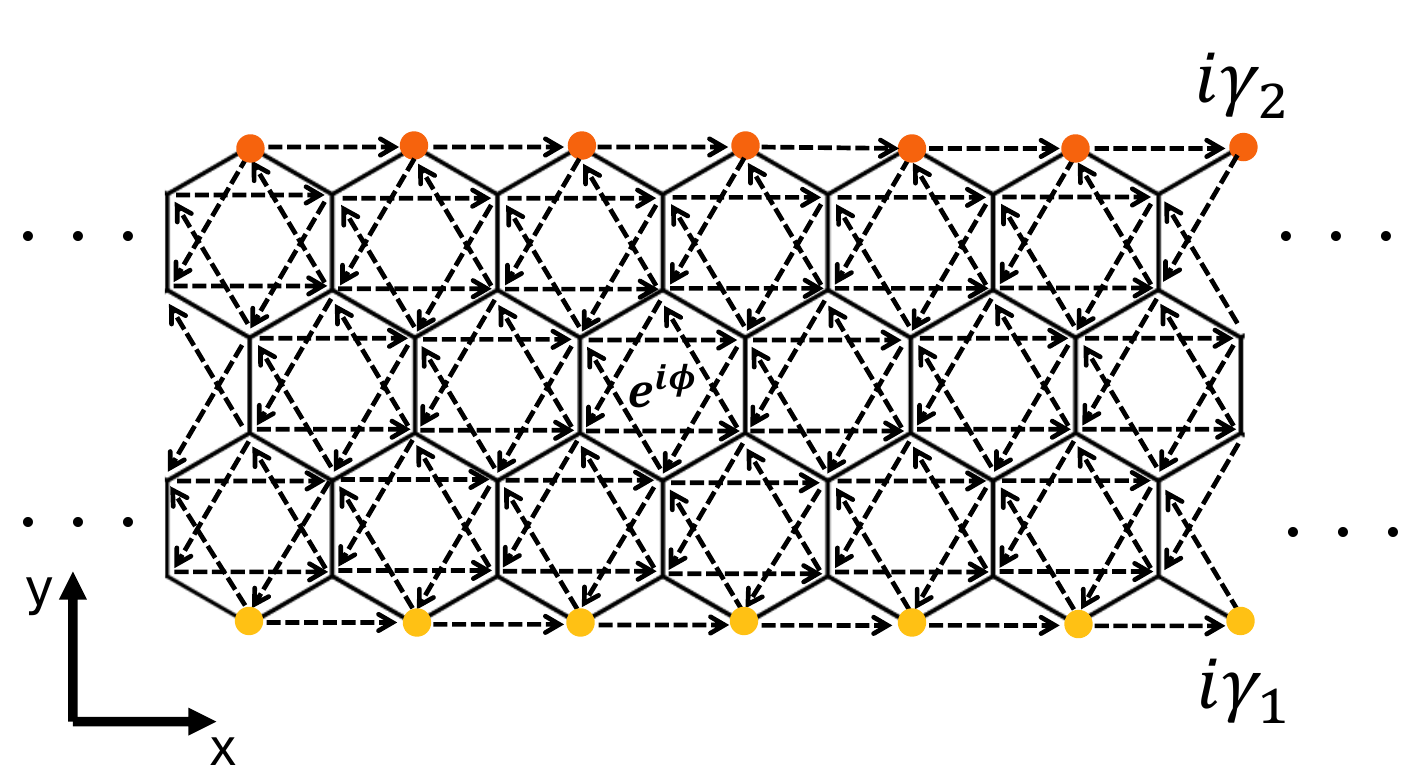}
    \caption{
    Schematic of the modified Haldane nanoribbon
    with edge-localized gain and loss.
     Imaginary on-site potentials $i\ga_1$ and $i\ga_2$ are applied to the lower and upper zigzag edges, respectively. The dashed arrows indicate the direction of phase $\phi$ in the NNN hopping.}
    \label{fig:mymodel}
\end{figure}
We consider a modified Haldane nanoribbon with imaginary on-site potentials introduced at the
two edges,  as illustrated in Fig.~\ref{fig:mymodel}.
Here we adopt the so-called zigzag edges, which host the antichiral edge states. 
Note that the antichiral edge states are absent on the armchair edges (left and right edges in Fig.~\ref{fig:mymodel}).
The lower and upper zigzag edges are
assigned imaginary on-site potentials
$i\ga_1$ and $i\ga_2$, respectively,
representing loss or gain depending on the sign of $\gamma_{1,2}$. The effective Hamiltonian is given by
\begin{equation}
    \begin{split}
        H(\ga_1,\ga_2)=&-t_1 \sum_{\langle i,j \rangle}c_i^\dagger c_j - t_2 \sum_{\langle\langle i,j \rangle\rangle}e^{i\nu_{ij}\phi}c_i^\dagger c_j  \\
        &+i\ga_1 \sum_{\substack{i\in \mathrm{L\text{-}edge}}}c_i^\dagger c_i+i\ga_2 \sum_{\substack{i\in \mathrm{U\text{-}edge}}}c_i^\dagger c_i.
    \end{split}
    \label{H}
\end{equation}
Here, $c_i^\dagger$ and $c_i$ denote the creation and annihilation operators at site $i$, respectively. The parameters $t_1$ and $t_2$ denote  the nearest-neighbor (NN) and NNN hopping amplitudes. The sums $\langle i,j \rangle$ and $\langle\langle i,j \rangle\rangle$ run over NN and NNN pairs, respectively. The factor $\nu_{ij}=\pm1$ specifies the direction of the phase $\phi$  acquired in the NNN hopping as indicated by the dashed arrows in Fig.~\ref{fig:mymodel}. 
Moreover, the sets $\mathrm{L\text{-}edge}$ ($\mathrm{U\text{-}edge}$) denote the lattice sites located on
the lower (upper) zigzag edge. 
\par
Let $N_x$ and $N_y$ be the numbers of lattice sites along the $x$ and $y$ directions, respectively. Because bulk states extend throughout the system, their wave-function weight on the gain/loss sites becomes negligible in the limit $N_y \to \infty$.
As a result, 
bulk states experience vanishing
effective non-Hermiticity and
 do not exhibit NHSE. In contrast, the antichiral edge states remain localized at the edges, and therefore, retain 
 sensitivity to edge gain and loss even for large $N_y$.

To elucidate how edge-localized gain and loss influence not only the edge states
but also the bulk spectrum, we focus on a narrow ribbon with small
$N_y$ and a large $N_x$. In this quasi-one-dimensional geometry, 
the edge dissipation significantly affects all eigenstates,
allowing both bulk and edge states to exhibit the NHSE. 
In addition,
we impose OBC along the $y$ direction throughout this work. Along the $x$ direction, we compare PBC and OBC
in order to identify the emergence of the NHSE.
We also perform calculations for a sufficiently wide ribbon, and the results are shown in Appendix~\ref{appenlargeNy}.

\subsection{\label{subsec:PTsymmetry}$\PT$ symmetry}
Under PBC along 
 the $x$ direction,
 the Hamiltonian~\eqref{H}
 satisfies 
\begin{equation}
    (\PT)H(\ga_1,\ga_2)(\PT)^{-1}=H(-\ga_2,-\ga_1),
    \label{symmetry}
\end{equation}
where $\mathcal{P}$ and $\mathcal{T}$ denote the inversion and  time-reversal operators, respectively. 
\par
\begin{figure}[t]
    \centering
    \includegraphics[width=0.9\linewidth]{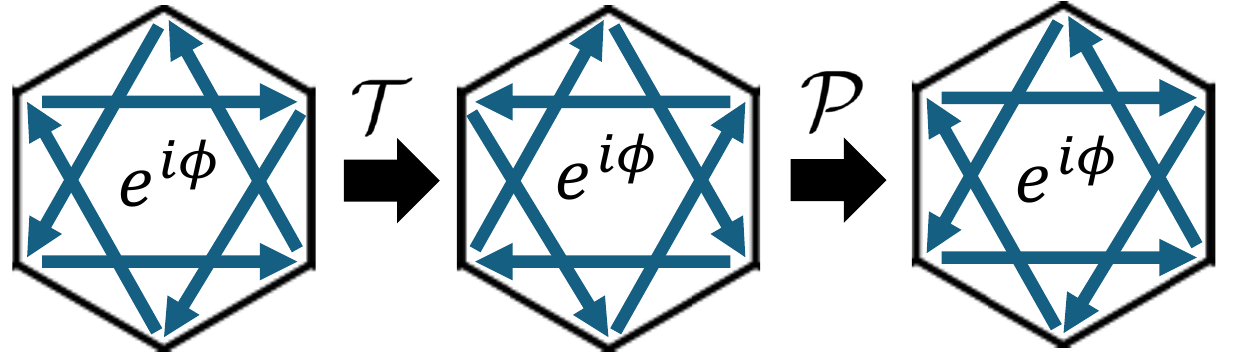}
    \caption{Illustration of the $\PT$ symmetry of the modified Haldane model. The direction of phase accumulation in the NNN hopping term is unchanged under the combined $\PT$ operation, ensuring the intrinsic $\PT$
    symmetry of the Hermitian part of the Hamiltonian. Here, $\mathcal{P}$ and $\mathcal{T}$ denote the inversion and time-reversal operators, respectively.}
    \label{fig:symmetry}
\end{figure}

This relation follows from  the intrinsic $\PT$ symmetry of the modified Haldane model 
(Fig.~\ref{fig:symmetry}). 
The  NNN hopping term in Eq.~\eqref{H} 
remains invariant under the combined $\PT$ operation.
In contrast, the imaginary on-site potentials
change sign under time reversal and are 
exchanged between the two edges under inversion.
As a result,
\begin{equation}
    \begin{split}
        &(\PT) \left( i\ga_1 \sum_{\substack{i\in \mathrm{L\text{-}edge}}}c_i^\dagger c_i +i\ga_2 \sum_{\substack{i\in \mathrm{U\text{-}edge}}}c_i^\dagger c_i \right) (\PT)^{-1} \\
        =& \mathcal{P}\left( -i\ga_1 \sum_{\substack{i\in \mathrm{L\text{-}edge}}}c_i^\dagger c_i -i\ga_2 \sum_{\substack{i\in \mathrm{U\text{-}edge}}}c_i^\dagger c_i \right) \mathcal{P}^{-1} \\
        =& -i\ga_2 \sum_{\substack{i\in \mathrm{L\text{-}edge}}}c_i^\dagger c_i -i\ga_1 \sum_{\substack{i\in \mathrm{U\text{-}edge}}}c_i^\dagger c_i.
    \end{split}
\end{equation}
Eq.~\eqref{symmetry} therefore
implies that the Hamiltonian possesses
 $\PT$ symmetry when $\ga_1=-\ga_2$;
 otherwise, the symmetry is explicitly broken.
\par
Let $E_{\gamma_1,\gamma_2}$ and $\psi_{\gamma_1,\gamma_2}(x,y)$ be an eigenenergy and eigenstate of $H(\gamma_1,\gamma_2)$. $(x,y)$ denotes the position of each lattice site. Then, we have
\begin{equation}
    H(\gamma_1,\gamma_2)\,\psi_{\gamma_1,\gamma_2}(x,y) =  E_{\gamma_1,\gamma_2}\,\psi_{\gamma_1,\gamma_2}(x,y).
    \label{shrodinger}
\end{equation}
Applying the $\PT$ operator 
yields
\begin{equation}
    \begin{split}
        &H(-\gamma_2,-\gamma_1)(\PT\psi_{\gamma_1,\gamma_2}(x,y)) \\
        =&E^*_{\gamma_1,\gamma_2}(\PT\psi_{\gamma_1,\gamma_2}(x,y)),
    \end{split}
    \label{conj}
\end{equation}
which shows that the spectra of 
$H(\gamma_1,\gamma_2)$ and $H(-\gamma_2,-\gamma_1)$ are complex conjugates of each other.

\section{\label{sec:HSTEanti}HSTE in antichiral edge states}
\begin{figure}
    \centering
    \includegraphics[width=0.99\linewidth]{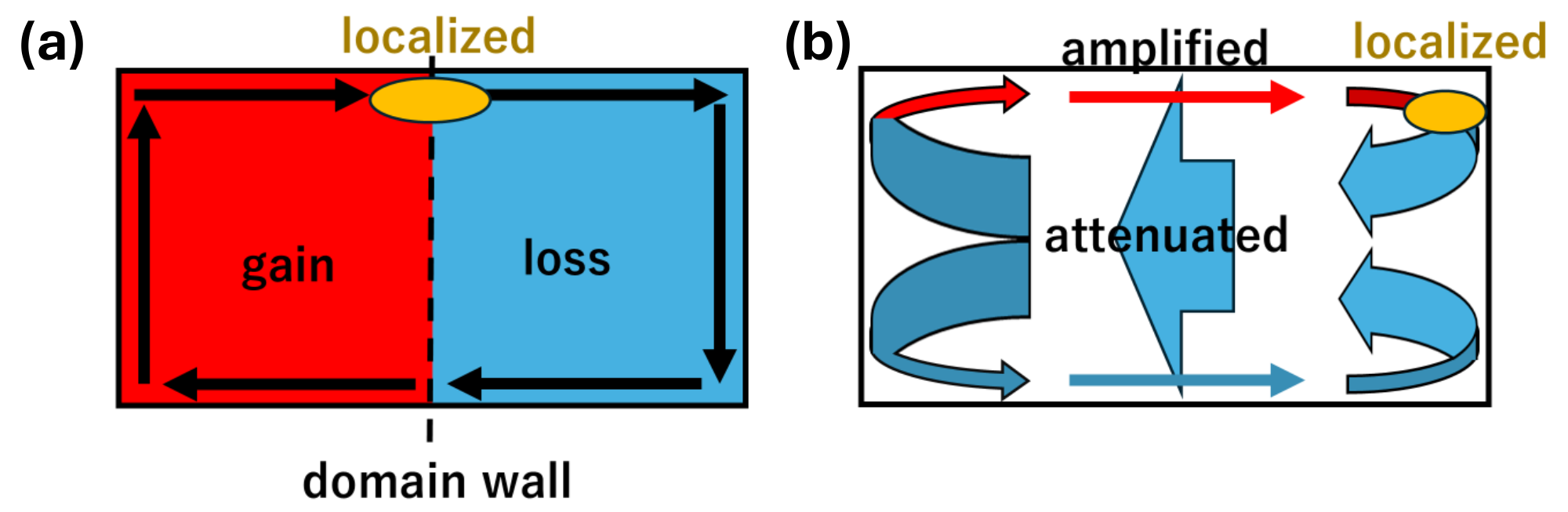}
    \caption{
    (a) HSTE of chiral edge states. The HSTE arises from the imbalance in amplification/attenuation between two connected chiral edge states. (b) HSTE of antichiral edge states. In contrast, the HSTE arises from the imbalance in amplification/attenuation between the edge states and the counterpropagating bulk states.
    }
    \label{fig:HSTEanti}
\end{figure}
Before presenting the numerical results, we describe the basic mechanism of the HSTE in antichiral systems
and contrast it with that in conventional chiral systems~\cite{2022PRL_HSTE,2022PRB_HSTE,Ma2024PRR,AnomalousHSTE}. 
The argument relies on  general properties of antichiral edge states: co-propagating edge states at opposite edges necessarily coexist with counterpropagating bulk states.
\par
In chiral systems, previous work has
shown that the chiral edge states are localized at the so-called domain wall of dissipation, i.e., the spatial interface where the imaginary potential profile changes~[Fig.~\ref{fig:HSTEanti}(a)]~\cite{Ma2024PRR}.  Since the edge states are amplified in the gain region and attenuated in the loss region, their amplitude reaches a maximum at the position where the system changes from amplification to attenuation. The HSTE arises from the interplay between two connected edge states.
\par
In antichiral systems, the situation is qualitatively different. 
For the propagation direction shown in~Fig.~\ref{fig:HSTEanti}(b), antichiral edge
states reach the right boundary under OBC along the $x$ direction. Because edge states are absent at the left and right boundaries of the system, these edge states are reflected at the right boundary and converted into counterpropagating bulk states.  The bulk states then
propagate back to the left, where they are reflected again and converted back into antichiral edge states. If the upper-edge states overlap with the gain sites, while the counterpropagating bulk states overlap with the loss sites, the upper-edge component is amplified whereas the counterpropagating bulk component is attenuated. Under OBC, these states localize at the position where the amplified states are converted into the attenuated states, namely, on the right side of the upper-edge states [Fig.~\ref{fig:HSTEanti}(b)]. Thus, the HSTE in antichiral edge states originates from the coupling between antichiral edge states and counterpropagating bulk states. This mechanism is distinct from the HSTE in conventional chiral edge states.
\par
To characterize whether a given eigenstate $\psi(x,y)$ under PBC is amplified or attenuated, we define the effective gain/loss of the eigenstate~\cite{Ma2024PRR}:
\begin{equation}
    \ga_{\mathrm{eff}}=\sum_{x,y}\ga(x,y)\,|\psi(x,y)|^2
    \label{eff}
\end{equation} 
which measures the net imaginary potential experienced by $\psi(x,y)$. In Eq.~\eqref{eff}, $\ga(x,y)$ denotes the imaginary part of the on-site potential at site $(x,y)$. A larger $\gamma_{\mathrm{eff}}$ corresponds to stronger amplification of $\psi(x,y)$, whereas a smaller $\gamma_{\mathrm{eff}}$ corresponds to stronger attenuation. For an antichiral edge state localized at the lower (upper) edge $\psi_{\mathrm{\mathrm{L\text{-}edge}}}(x,y)$ ($\psi_{\mathrm{\mathrm{U\text{-}edge}}}(x,y)$), the effective gain/loss is denoted by $\gamma_{\mathrm{eff}}^{\mathrm{L\text{-}edge}}$ $\big(\gamma_{\mathrm{eff}}^{\mathrm{U\text{-}edge}}\big)$.
For a bulk state $\psi_{\mathrm{bulk}}(x,y)$, the effective gain/loss is denoted by $\gamma_{\mathrm{eff}}^{\mathrm{Bulk}}$. Because a uniform shift of the imaginary part of the energy leaves the eigenstates of the Hamiltonian unchanged, the effective gain/loss is determined relative to a reference value rather than on an absolute scale. 
\par
When $\gamma_{\mathrm{eff}}^{\mathrm{U\text{-}edge}}>\gamma_{\mathrm{eff}}^{\mathrm{Bulk}}$, the upper-edge states are relatively amplified as they propagate to the right, whereas the counterpropagating bulk states are relatively attenuated as they propagate to the left. Under OBC, these states become hybridized and accumulate on the right side of the upper edge, as shown in Fig.~\ref{fig:HSTEanti}(b). When $\gamma_{\mathrm{eff}}^{\mathrm{U\text{-}edge}}<\gamma_{\mathrm{eff}}^{\mathrm{Bulk}}$, the situation is reversed, and the states become localized at the left side of the upper edge.
\par
We reveal that applying loss exclusively to the lower boundary induces localization of antichiral skin modes at the upper edge. This implies a phenomenon that we call the \textit{nonlocal antichiral skin effect}, where the antichiral skin modes localize away from the lossy sites. This phenomenon can be explained as follows. Since the bulk states are spatially extended over the entire system, they are affected by the lossy sites at the lower edge. As a result, they are attenuated by these sites, leading to $\gamma_{\mathrm{eff}}^{\mathrm{Bulk}}<\gamma_{\mathrm{eff}}^{\mathrm{U\text{-}edge}}$. Consequently, the situation shown in Fig.~\ref{fig:HSTEanti}(b) is realized. This phenomenon originates from the distinctive feature of antichiral systems: the presence of counterpropagating bulk modes. We analyze this effect in detail in the next section.

\section{\label{sec:result}result}
Throughout this section, we fix $t_1=1$, $t_2=0.05$, $\phi=\pi/2$, and $N_y=8$.
We investigate three
representative parameter sets:
\begin{enumerate}[label=(\roman*)]
    \item $\ga_1=-0.6,\ga_2=+0.6$,
    \item $\ga_1=-0.6,\ga_2=0$,
    \item $\ga_1=0,\ga_2=+0.6$.
\end{enumerate}
Case (i) satisfies the $\PT$-symmetry condition.
In this regime, only the antichiral edge 
states exhibit the NHSE, while the bulk
states remain extended. In contrast,
cases (ii) and (iii) explicitly break $\PT$ symmetry,
allowing both edge and bulk states to develop
the NHSE. The latter two cases are related to each
other by the $\PT$ transformation.

\subsection{\label{subsec:LU}Lower-edge loss and upper-edge gain}
\begin{figure}[tb]
    \centering
    \includegraphics[width=0.99\linewidth]{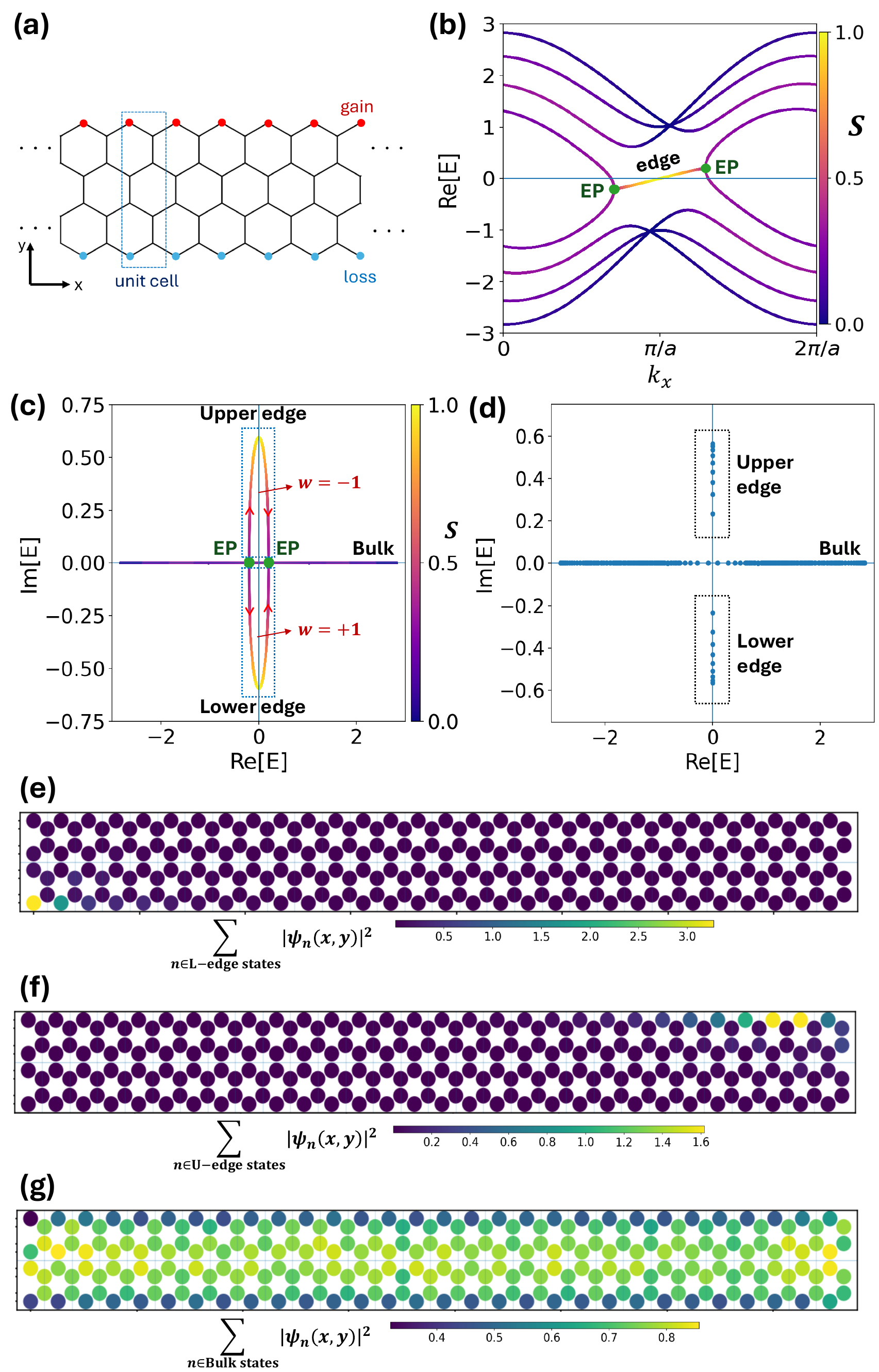}
    \caption{Results for the $\PT$-symmetric case $H(\ga_1=-0.6,\ga_2=+0.6)$. (a) Schematic of the system. (b) Band structure of $\mathrm{Re}[E]$ under PBC along the $x$ direction. (c) Complex-energy spectrum under PBC. The red arrows indicate the direction of each loop as $k_x$ increases, and $w$ denotes the point-gap winding number. In (b) and (c), brighter colors indicate stronger localization at the upper and lower edges. (d) Energy spectrum under OBC along $x$. (e) Spatial distribution of the summed lower-edge states enclosed by the dashed rectangle in (d). (f) Spatial distribution of the summed upper-edge states enclosed by the dashed rectangle in (d). (g) Spatial distribution of the summed bulk states on the real axis in (d). In (e)-(g), brighter colors indicate higher state density. In (d)-(g), the system size is $N_x=30$.
    }
    \label{fig:UgainLloss}
\end{figure}

We first consider the case $H(\ga_1=-0.6,\ga_2=+0.6)$, shown in Fig.~\ref{fig:UgainLloss}(a). According to Eq.~\eqref{symmetry}, this system satisfies the $\PT$-symmetry condition $\gamma_1=-\gamma_2$.
\subsubsection{Spectrum under PBC}
Under PBC along the $x$ direction, 
the Bloch wave number $k_x$ is well defined.
The real part of the band structure and the complex-energy spectrum are shown in Figs.~\ref{fig:UgainLloss}(b) and ~\ref{fig:UgainLloss}(c), respectively. 
To characterize edge localization, we define
\begin{equation}
    S=\frac{\sum\limits_x|\psi(x,y_{\min})|^2+\sum\limits_x|\psi(x,y_{\max})|^2}{\sum\limits_{x,y} |\psi(x,y)|^2},
    \label{S}
\end{equation}
where $y_{\min}$ and $y_{\max}$ denote the lower and upper edges, respectively. 
In Figs.~\ref{fig:UgainLloss}(b) and \ref{fig:UgainLloss}(c), brighter colors indicate larger values of $S$, corresponding to stronger localization at the upper and lower edges.
\par
As seen in Fig. 5(b), a pair of exceptional points (EPs) emerges where the
antichiral edge branch merges into the bulk continuum. These EPs lie close to the Dirac points in Fig.~\ref{fig:anti}(c), but they do not coincide exactly with them, because the ribbon
considered here is narrow. As indicated by the color of the bands, the states lying between the two EPs
are the antichiral edge states.
From Fig.~\ref{fig:UgainLloss}(c), we find that 
all bulk eigenenergies lie on the real axis,
whereas the eigenenergies of antichiral edge states form
complex-conjugate pairs.
This indicates that 
all bulk states belong to the $\PT$-symmetric phase,
\begin{equation}
    \PT\,\psi_{\mathrm{bulk}}(x,y)=\psi^*_{\mathrm{bulk}}(-x,-y)=\psi_{\mathrm{bulk}}(x,y),
    \label{bulkPT}
\end{equation}
whereas all upper-edge states and lower-edge states are related by the $\PT$ transformation, 
\begin{equation}
    \PT\,\psi_{\mathrm{L\text{-}edge}}(x,y)=\psi^*_{\mathrm{L\text{-}edge}}(-x,-y)=\psi_{\mathrm{U\text{-}edge}}(x,y).
\end{equation}
\par
In Fig.~\ref{fig:UgainLloss}(b), since the group velocity along the $x$ direction is given by
\begin{equation}
    v_{x}=\frac{1}{\hbar}\frac{\partial\,\mathrm{Re}[E]}{\partial k_x},
    \label{v}
\end{equation}
the antichiral edge states on both edges propagate in the same direction, consistent with Fig.~\ref{fig:anti}(b). As the lower-edge states experience only loss sites, whereas the upper-edge states experience only gain sites, the signs of $\mathrm{Im}[E]$ are opposite.
\par
In the complex-energy plane [Fig.~\ref{fig:UgainLloss}(c)], the upper- and lower-edge states each form a distinct loop with the bulk states located between the two EPs. The red arrows in Fig.~\ref{fig:UgainLloss}(c) indicate the direction of each loop as $k_x$ increases. The winding number
\begin{equation}
    w(E_b)=\frac{1}{2\pi i}\int_{0}^{2\pi} \mathrm{d}k_x \,
    \frac{\partial}{\partial k_x}\log\det\!\left[H(k_x)-E_b\right]
    \label{winding}
\end{equation}
is $+1$ for the lower-edge loop and $-1$ for the upper-edge loop. 
In Eq.~\eqref{winding}, $E_b$ denotes the base energy, which is chosen to lie inside the loop whose winding number is evaluated. The nonzero winding numbers indicate that the antichiral edge states exhibit the NHSE. 
\par
In contrast, all bulk states, except those forming loops together with the edge states [i.e., the bulk states lying between the two EPs in Fig. 5(c)], do not have a point gap and therefore have zero winding number. As a result,  these bulk states do not exhibit the NHSE under OBC. The extended nature of them is protected by the $\PT$ symmetry because all of their eigenenergies lie on the real axis and therefore cannot open a point gap. 

\subsubsection{Spectrum under OBC}
We now impose OBC along the $x$ direction.
In this case, the two loops observed under PBC collapse
into line segments, as
shown in Fig.~\ref{fig:UgainLloss}(d)
\footnote{
To clarify the correspondence between the spectra under PBC and OBC, we continuously
deform the boundary condition by multiplying the hopping terms connecting the left and right boundaries by a parameter $\lambda$,
varying it from  $\lambda=1$ (PBC) to $\lambda=0$ (OBC).
}.
\par
Under OBC, the lower-edge states localize at 
the left boundary, whereas the upper-edge 
states localize at the right boundary
[Fig.~\ref{fig:UgainLloss}(e) and Fig.~\ref{fig:UgainLloss}(f)].
The localization direction is consistent with
the sign of the winding number of each loop calculated in Eq.~\eqref{winding}. This is an HSTE induced by antichiral edge states. The physical mechanism of this phenomenon is described in Sec.~\ref{sec:HSTEanti}. 
That is, under OBC the antichiral edge modes hybridize with the
counterpropagating bulk modes and turn into antichiral skin modes. The bulk
modes involved are those forming the loops with the edge modes in Fig. 5(c),
i.e., the bulk states lying between the two EPs.
\par
In contrast, consistent with the winding numbers under PBC, the other bulk states remain extended under OBC [Fig.~\ref{fig:UgainLloss}(g)]. 
This phenomenon can also be understood from Eq.~\eqref{bulkPT}. 
In the $\PT$-symmetric regime, the condition
\begin{equation}
    |\psi_{\mathrm{bulk}}(x,y)|^2 = |\psi_{\mathrm{bulk}}(-x,-y)|^2,
\end{equation}
holds, preventing spatially asymmetric accumulation.
Thus, $\PT$ symmetry protects the bulk
states from exhibiting the NHSE.
\par
To be precise, under OBC, the shapes of the left and right edges are slightly different, as shown in Fig.~\ref{fig:UgainLloss}(a). 
Thus, $\PT$ symmetry is
 broken by the finite-size effect. 
When $N_x$ is sufficiently large, however, the effects of the left and right edges are negligible and have little influence on the presence or absence of the skin effect.

\subsubsection{Critical strength of gain/loss}
By increasing  $|\ga_1|=|\ga_2|$,
we find a critical value $|\ga_1|=|\ga_2|>\ga_0\approx0.89$ above which 
bulk states enter the $\PT$-broken phase. 
In the main text, we restrict ourselves to the  regime $|\ga_{1,2}|<\ga_0$, where all bulk states remain $\PT$ symmetric. The $\PT$-broken regime is discussed in Appendix~\ref{appenA}.

\subsection{\label{subsec:L}Lower-edge loss}
\begin{figure*}[tb]
    \centering
    \includegraphics[width=0.99\linewidth]{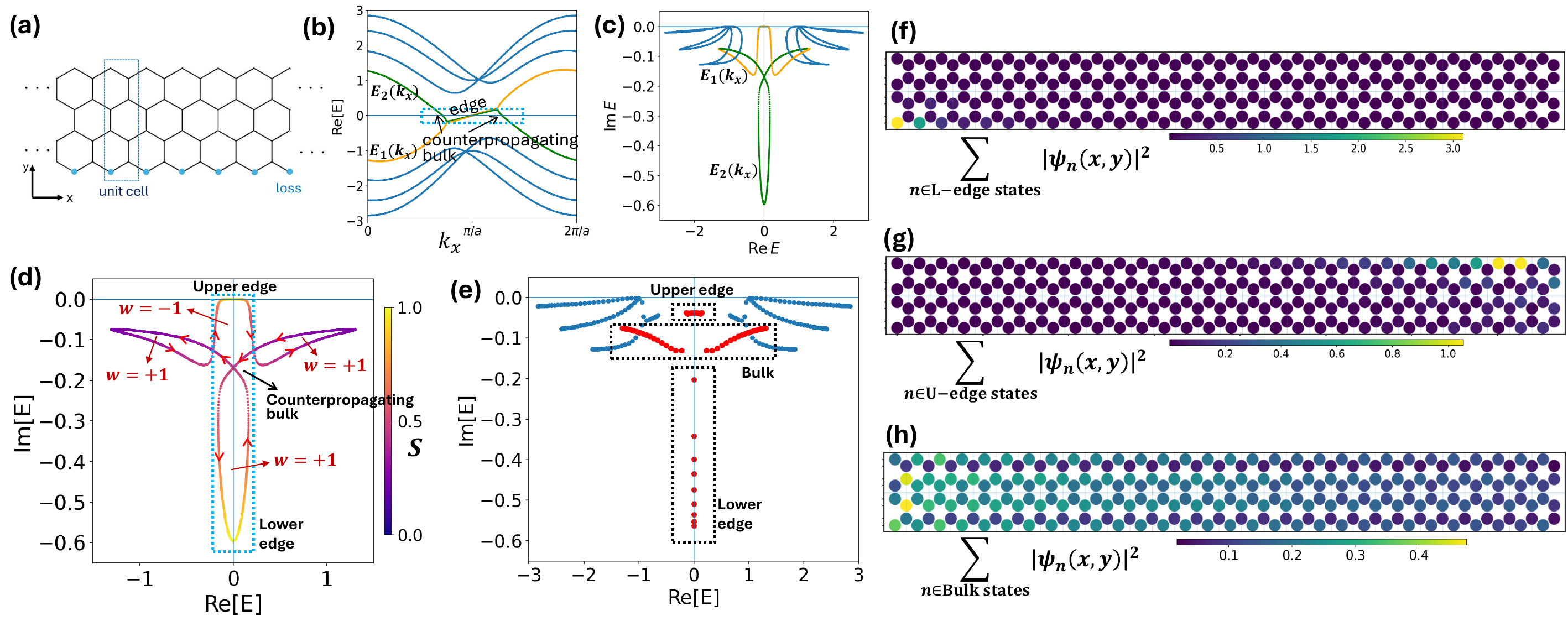}
    \caption{Results for $H(\ga_1=-0.6,\ga_2=0)$. (a) Schematic of the system. (b) Band structure of $\mathrm{Re}[E]$ under PBC along the $x$ direction. In the main text, we focus on the bands $E_1$ and $E_2$. (c) Complex-energy spectrum under PBC. (d) Complex-energy spectrum of $E_1$ and $E_2$ under PBC. The red arrows indicate the direction of each loop as $k_x$ increases, and $w$ denotes the point-gap winding number. Brighter colors indicate stronger localization at the upper and lower edges. The eigenenergies enclosed by the blue dashed rectangle correspond to those enclosed in (b). (e) Energy spectrum under OBC along $x$. Red dots indicate eigenenergies originating from $E_1$ and $E_2$ in (c). (f) Spatial distribution of the summed lower-edge states enclosed by the dashed rectangle in (e). (g) Spatial distribution of the summed upper-edge states enclosed by the dashed rectangle in (e). (h) Spatial distribution of the summed bulk states (red dots) enclosed by the dashed rectangle in (e). In (f)-(h), brighter colors indicate higher state density. In (e)-(h), the system size is $N_x=30$.
    }
    \label{fig:Lloss}
\end{figure*}

We next consider the case $H(\ga_1 = -0.6, \ga_2 = 0)$,
shown in Fig. 6(a), where $\PT$ symmetry is explicitly broken.
\par
Under PBC, the band structure of $\mathrm{Re}[E]$ and the energy spectrum are shown in Figs.~\ref{fig:Lloss}(b) and \ref{fig:Lloss}(c), respectively. 
In the main text, we focus on the band containing all antichiral edge states and some of the bulk states, which appears as two branches, $E_1(k_x)$ and $E_2(k_x)$, shown in yellow and green in Fig.~\ref{fig:Lloss}(b), respectively. 
There are no band degeneracies between $E_1(k_x)$ and $E_2(k_x)$, nor between either of these two bands and the blue bands
\footnote{
For $N_y=8$, which is fixed throughout this study, each band evolves 
independently under OBC without mixing with other bands.
We point out that for larger $N_y$, bulk states from different bands may hybridize under OBC,  complicating the spectral evolution. 
}.
$E_1$ and $E_2$ form four loops
in the complex-energy plane, as shown in Fig.~\ref{fig:Lloss}(c), indicating that all eigenstates in this band exhibit the NHSE under OBC. Other bands also form loops and 
 therefore exhibit the NHSE; their behavior is  discussed in Appendix~\ref{appenB}.
\par
We extract $E_1(k_x)$ and $E_2(k_x)$ and show their spectrum separately in Fig.~\ref{fig:Lloss}(d). 
The color scale represents the quantity $S$ defined in Eq.~\eqref{S}, which characterizes the degree of edge localization. The red arrows indicate the direction of evolution as $k_x$ increases. The spectrum can be divided into two parts.

\subsubsection{NHSE in antichiral edge states}
The first part, enclosed by the blue dashed rectangle in Fig.~\ref{fig:Lloss}(d),  corresponds to the same region in Fig.~\ref{fig:Lloss}(b). 
From Eq.~\eqref{v}, 
this region includes
 all antichiral edge states propagating in the $+x$ direction, together with bulk states propagating in the $-x$ direction. 
For instance, at $\mathrm{Re}[E]=0$, two antichiral edge states propagate to the right while two bulk states propagate to the left.
In Fig.~\ref{fig:Lloss}(d), this region forms two loops: one loop contains all lower-edge states and some counterpropagating bulk states with the winding number $w=+1$, while the other contains all upper-edge states and the remaining counterpropagating bulk states with $w=-1$.
\par
Under OBC, these two loops collapse into two line segments [Fig.~\ref{fig:Lloss}(e)], indicating hybridization between the edge states and the counterpropagating bulk states. The physical mechanism is described in detail in Sec.~\ref{sec:HSTEanti}. The lower-edge states localize on
the left side
[Fig.~\ref{fig:Lloss}(f)], while the upper-edge
states localize on the right side [Fig.~\ref{fig:Lloss}(g)]. 
\par
It is important to note that no gain or loss is applied to the upper edge. 
Nevertheless, the NHSE also emerges at the upper-edge states.
Since the localized position (the right side of the upper edge) is far from the domain wall of dissipation (between the lower edge and the bulk),
we call this phenomenon the
\textit{nonlocal antichiral skin effect}.
The physical mechanism of this phenomenon is as described in Sec.~\ref{sec:HSTEanti}.
Since the counterpropagating bulk states extend over the entire system, they are attenuated by the loss introduced at the lower edge.  In contrast, since the upper-edge states are far from the lossy sites at the lower edge, they are not attenuated. As a result, $\gamma_{\mathrm{eff}}^{\mathrm{Bulk}}<\gamma_{\mathrm{eff}}^{\mathrm{U\text{-}edge}}\approx0$, so that the upper-edge states are relatively amplified, realizing the situation shown in Fig.~\ref{fig:HSTEanti}(b).
\par
Moreover, if the loss at the lower edge is replaced by gain, that is, $\ga_{\mathrm{eff}}^{\mathrm{Bulk}}>\ga_{\mathrm{eff}}^{\mathrm{U\text{-}edge}}$ is realized, the upper-edge states become localized on the left side. Therefore, in this model, the edge skin effect can be controlled by gain and loss applied at spatially separated locations.
\par
Note that in the limit $N_y \to \infty$,
the loss at the lower edge has little effect on the bulk states, leading to  
$\ga_{\mathrm{eff}}^{\mathrm{Bulk}}  \to 0$.
In this limit, the NHSE of the upper-edge states disappears, while that of the lower-edge states persists. 
However, by enhancing the loss, it is possible to induce the NHSE of the upper-edge states even in sufficiently wide systems. The numerical results for sufficiently large $N_y$ are shown in Appendix~\ref{appenlargeNy}.
\par
Because the antichiral edge states appear only along the upper and lower boundaries, the number of edge states is proportional to the number of lower- and upper-boundary sites. Consequently, the number of antichiral skin modes is proportional to $N_x$, but is independent of $N_y$. 

\subsubsection{NHSE in bulk states}
The second part of the spectrum in Fig.~\ref{fig:Lloss}(d),
located outside the blue dashed rectangle, 
consists purely of bulk states with both propagation directions. These states also form  two loops,  both with winding numbers  $w=+1$, indicating that they localize in the same direction under OBC. The bulk NHSE  is not forbidden here, since the system lacks $\PT$ symmetry. These bulk states localize on the left side under OBC [Fig.~\ref{fig:Lloss}(h)]. These NHSEs are also induced by the imbalance in spatial flow caused by $\phi$.
\par
The spatial distribution of each eigenstate in the band diagram, together
with a more detailed physical interpretation of the NHSE, is provided in
Appendix~\ref{appen:NHSEinter}.

\subsection{\label{subsec:U}Upper-edge gain}
\begin{figure}
    \centering
    \includegraphics[width=0.99\linewidth]{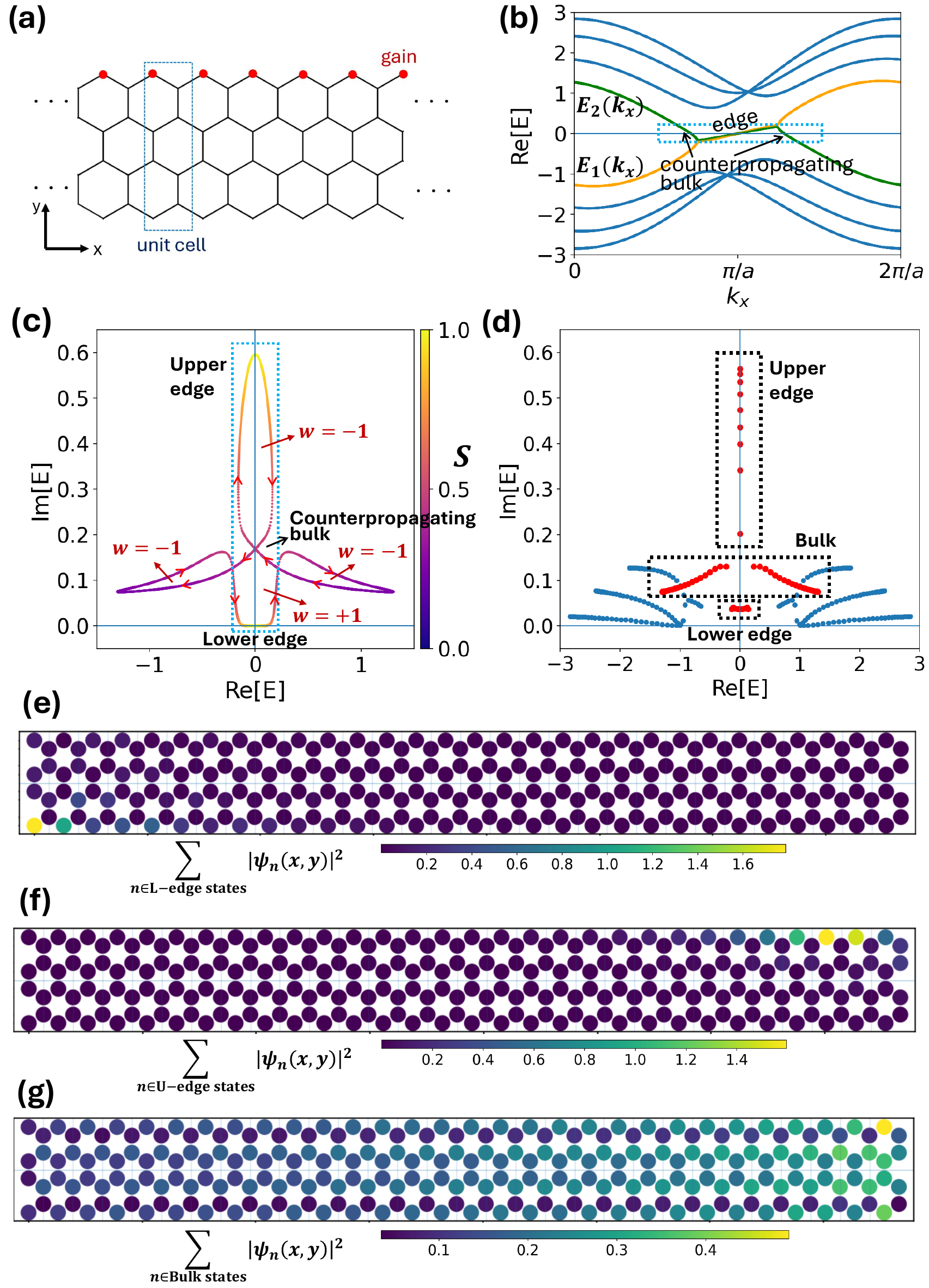}
    \caption{Results for $H(\ga_1=0,\ga_2=+0.6)$,
    the $\PT$-counterpart of Fig.~\ref{fig:Lloss}. (a) Schematic of the system. (b) Band structure of $\mathrm{Re}[E]$ under PBC along the $x$ direction. 
    (c) Complex-energy spectrum of $E_1$ and $E_2$ under PBC. The red arrows indicate the direction of each loop as $k_x$ increases, and $w$ denotes the point-gap winding number. 
    (d) Energy spectrum under OBC along $x$. 
    (e) Spatial distribution of the summed lower-edge states enclosed by the dashed rectangle in (d). 
    (f) Spatial distribution of the summed upper-edge states enclosed by the dashed rectangle in (d). 
    (g) Spatial distribution of the summed bulk states  enclosed by the dashed rectangle in (d).
    In (e)-(g), brighter colors indicate higher state density. In (d)-(g), the system size is $N_x=30$.
    Panels (c)-(g) correspond to Figs.~\ref{fig:Lloss}(d)-\ref{fig:Lloss}(h), but with opposite winding numbers
    due to the $\PT$ transformation.
    }
    \label{fig:Ugain}
\end{figure}
In this subsection, we consider the Hamiltonian $H(\ga_1=0,\ga_2=+0.6)$, shown in Fig.~\ref{fig:Ugain}(a). According to Eq.~\eqref{symmetry}, this system is the $\PT$ counterpart of $H(\ga_1=-0.6,\ga_2=0)$, namely,
\begin{equation}
    \begin{split}
        &(\PT)H(\ga_1=0,\ga_2=+0.6)(\PT)^{-1}\\
        =&H(\ga_1=-0.6,\ga_2=0).
    \end{split}
\end{equation}
From Eq.~\eqref{conj}, it follows that the eigenenergies of $H(\ga_1=0,\ga_2=+0.6)$ are the complex conjugates of those of $H(\ga_1=-0.6,\ga_2=0)$, and the corresponding eigenstates are related by the $\PT$ operation.
\par
Figure~\ref{fig:Ugain}(b) shows the band structure under PBC. 
As in the previous subsection, we focus on the two bands $E_1(k_x)$ and $E_2(k_x)$, whose complex-energy spectrum is shown in Fig.~\ref{fig:Ugain}(c). This spectrum is the complex conjugate of Fig.~\ref{fig:Lloss}(d);
accordingly, all winding numbers change sign.
The energy spectrum under OBC is shown in
Fig.~\ref{fig:Ugain}(d), which is likewise the complex conjugate of Fig.~\ref{fig:Lloss}(e). 
\par
Figures~\ref{fig:Ugain}(e)--(g) display the spatial distributions of the lower-edge, upper-edge, and bulk states corresponding to the red points in Fig.~\ref{fig:Ugain}(d),  originating from $E_1$ and $E_2$.
In principle, these states are related to
those in Fig.~\ref{fig:Lloss} by the $\PT$
transformation.
However, since 
the left and right boundaries are not
strictly identical under OBC, 
the $\PT$ mapping is not exact in the finite system. Nevertheless, the direction of the NHSE is  consistent with 
the $\PT$ mapping.
\par
A detailed physical interpretation of the NHSE is provided in Appendix~\ref{appen:NHSEinter}.

\subsection{\label{subsec:phase}Phase diagram of the bulk NHSE}
In this subsection, we show that the NHSE of the bulk states does not arise simply from the ambiguity between bulk and edge states in a finite-size system. The bulk states are affected by non-Hermiticity in a manner distinct from the edge states. 
\par
In Figs.~\ref{fig:UgainLloss}, \ref{fig:Lloss}, and \ref{fig:Ugain}, we present the spatial profiles of antichiral edge states and bulk states under OBC in three cases. 
In all these cases, we 
observe that the upper-edge states are localized on the right boundary, while the lower-edge states are localized on the left boundary. In contrast, the bulk states originating from $E_1$ and $E_2$ remain extended in the $\PT$-symmetric case ($\ga_1=-0.6,\:\ga_2=+0.6$) [Fig.~\ref{fig:UgainLloss}(g)], localize on the left side for ($\ga_1=-0.6,\:\ga_2=0$) [Fig.~\ref{fig:Lloss}(h)], and localize on the right side for ($\ga_1=0,\:\ga_2=+0.6$) [Fig.~\ref{fig:Ugain}(g)].
\par
\begin{figure}[tb]
    \centering
    \includegraphics[width=0.7\linewidth]{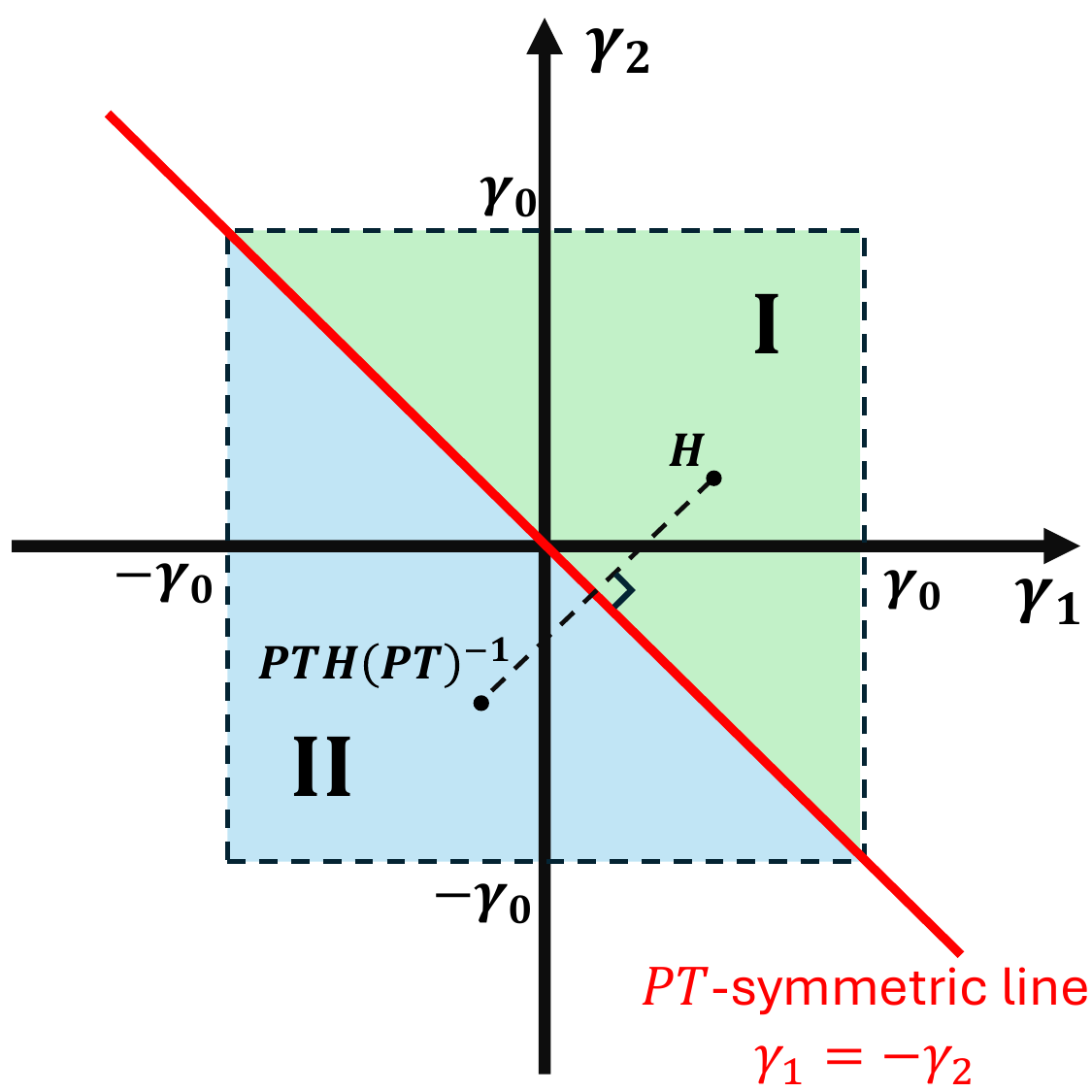}
    \caption{Phase diagram of the NHSE in the $(\gamma_1,\gamma_2)$ plane. The red line $\gamma_1=-\gamma_2$ corresponds to the $\PT$-symmetric regime, where the bulk NHSE is forbidden. 
    For $|\gamma_1|=|\gamma_2|<\gamma_0$, all bulk
    states remain in the $\PT$-symmetric phase.
    The $\PT$-symmetric line separates Phase I ($\gamma_1+\gamma_2>0$) and Phase II ($\gamma_1+\gamma_2<0$), in which the bulk skin modes
    localize at opposite boundaries. The two phases are
    related by the $\PT$ transformation.
    }
    \label{fig:phase}
\end{figure}
We find that the localization direction of the bulk states is determined solely by the sign of $\ga_1+\ga_2$, 
 and summarize these results in the phase diagram shown in Fig.~\ref{fig:phase}.
From Eq.~\eqref{symmetry},  the Hamiltonian $H(\gamma_1,\gamma_2)$ possesses $\PT$ symmetry along the line $\gamma_1=-\gamma_2$. As discussed in Sec.~\ref{subsec:LU}, there exists a critical value $\gamma_0 \approx 0.89$. When $|\gamma_1|=|\gamma_2|<\gamma_0$, all bulk states remain in the $\PT$-symmetric phase, whereas the antichiral edge states belong to the $\PT$-broken phase. In what follows, we restrict our attention to the regime $|\gamma_{1,2}|<\gamma_0$.
\par
The $\PT$-symmetric line $\gamma_1+\gamma_2=0$ divides this region into two parts, denoted as Phase~I and Phase~II. According to Eq.~\eqref{symmetry},  the Hamiltonians in these two phases are related by the $\PT$ transformation. 
\par
In Phase~I ($\gamma_1+\gamma_2>0$), the two edges provide
a net gain to the system. Since bulk states extend over both edges, their eigenenergies 
 satisfy $\mathrm{Im}[E]\ge0$. 
 Under OBC, the bulk states originating from $E_1$ and $E_2$ localize on the right boundary. 
 $H(\ga_1=0,\ga_2=+0.6)$ belongs to phase I.
\par
In Phase~II ($\gamma_1+\gamma_2<0$), the situation is reversed: the system has a net loss and the bulk eigenenergies satisfy $\mathrm{Im}[E]\le0$. 
Under OBC, the bulk states originating from $E_1$ and $E_2$  localize on the left boundary. 
The localization direction is opposite to that in phase I because all eigenenergies are mapped to their complex conjugates under the $\PT$ transformation, and the winding numbers of the point gaps are reversed.
$H(\ga_1=-0.6,\ga_2=0)$ belongs to phase II.
\par
The physical mechanism by which the bulk states exhibit the NHSE is discussed in Appendix~\ref{appen:NHSEinter}. The NHSE of bulk states belonging to bands other than $E_1$ and $E_2$ is discussed in Appendix~\ref{appenB}.
\par
Note that the NHSE of the bulk states disappears when the system size along the $y$ direction is sufficiently large, because for large $N_y$ the gain and loss sites at the edge have little influence on the bulk states. It follows that the NHSE of the bulk states originates from the finite width of the system. The numerical results for sufficiently large $N_y$ are shown in Appendix~\ref{appenlargeNy}.
\par
Finally, we comment on the antichiral edge states.
Unlike the bulk states, their localization direction
cannot be determined solely from the phase diagram in Fig.~\ref{fig:phase}. 
The localization of the upper (lower) edge states depends
on the relative magnitude of the effective gain/loss of 
the edge states, $\gamma_{\mathrm{eff}}^{\text{edge}}$,
and that of the counterpropagating bulk states, 
$\gamma_{\mathrm{eff}}^{\text{bulk}}$.
Because $\gamma_{\mathrm{eff}}^{\text{bulk}}$ depends on
the detailed spatial distribution of the bulk states,
the edge-state localization direction cannot be 
expressed in a simple global phase diagram.

\section{\label{conclusion}conclusion}
In this work, we have investigated a narrow non-Hermitian modified Haldane nanoribbon with gain and loss applied at the zigzag edges, and analyzed the non-Hermitian skin effect (NHSE) in both antichiral edge states and bulk states. 
\par
First, we clarified the mechanism underlying the hybrid skin-topological effect (HSTE) in the antichiral systems. In contrast to the conventional HSTE in chiral systems, where the NHSE originates from the interplay of edge modes at the domain wall of dissipation, the HSTE in the present antichiral system arises from the competition between the antichiral edge states and the counterpropagating bulk states. The localization direction of the antichiral edge states is determined by the relative magnitude of the effective gain/loss experienced by the edge states and by the counterpropagating bulk states. 
\par
Second, we established the role of the $\PT$ symmetry 
in organizing the bulk NHSE. 
In the $\PT$-symmetric regime, the bulk eigenenergies
remain real and no point gap opens, thereby forbidding the bulk NHSE.  
Once $\PT$ symmetry is broken by tuning the edge gain and loss, 
the bulk NHSE emerges. 
The $\PT$-symmetric line separates two distinct bulk-skin
phases characterized by opposite localization directions.
\par
Our results demonstrate that edge-localized dissipation
can serve as a symmetry-based control mechanism for
non-Hermitian skin effects in the modified Haldane model.
More broadly, this work highlights the qualitative
differences between chiral and antichiral topological structures in non-Hermitian settings and provides a framework for engineering controllable skin modes via boundary design.
Our model may be experimentally realizable in synthetic platforms such as photonic crystals or topolectrical circuits, where gain and loss can be engineered in a controlled manner.

\section*{acknowledgment}
We are grateful to K. Kakimoto for helpful discussions and valuable comments on the manuscript. 
SU is supported by JST PRESTO~(JPMJPR2351) and
JSPS KAKENHI~(JP25K07191).

\appendix
\section{\label{appenA}$\PT$-symmetric system with large gain and loss}
\begin{figure}[!htb]
    \centering
    \includegraphics[width=0.99\linewidth]{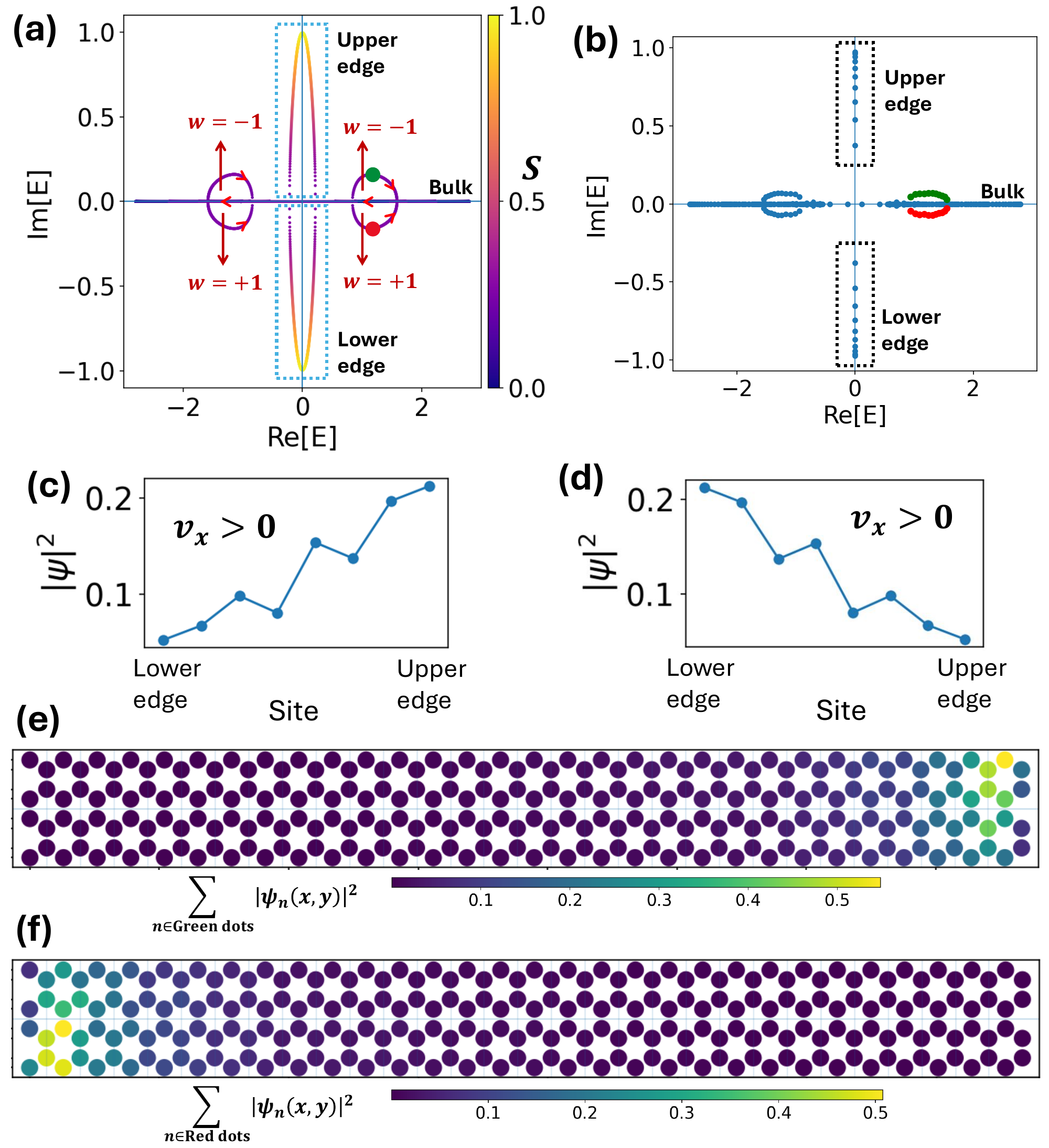}
    \caption{Results for $H(\ga_1=-1,\ga_2=+1)$,
    where $|\gamma_1|=|\gamma_2|>\gamma_0$. (a) Complex-energy spectrum under PBC along the $x$ direction.
    The red arrows indicate the direction of each loop as $k_x$ increases, and $w$ denotes the point-gap winding number. Brighter colors indicate stronger localization at the upper and lower edges. 
    (b) Energy spectrum under OBC along $x$. 
    (c) Spatial profiles of the green dot in (a). 
    (d) Spatial profiles of the red dot in (a). 
    In (c) and (d), the displayed region
    corresponds to the sites within the unit cell shown in Fig.~\ref{fig:UgainLloss}(a).
    $v_x$ denotes the group velocity along the $x$ direction of the state shown.
    (e) Spatial distribution of the summed bulk states corresponding to the green points in (b). 
    (f) Spatial distribution of the summed bulk states corresponding to the red points in (b). 
    In (e) and (f), brighter colors indicate higher state density. In (b), (e) and (f), the system size is $N_x=30$.
    }
    \label{fig:appenA}
\end{figure}
In this Appendix, we present the results for the Hamiltonian $H(\ga_1=-1,\ga_2=+1)$. In this case, $|\ga_1|=|\ga_2|>\ga_0\approx 0.89$, and some bulk states enter the $\PT$-broken phase.
\par
Figure~\ref{fig:appenA}(a) shows the complex-energy spectrum under PBC. 
The spatial distributions within a unit cell
for the representative green and red points are shown 
in Fig.~\ref{fig:appenA}(c) and \ref{fig:appenA}(d), respectively. These results indicate that bulk states in the $\PT$-broken phase become spatially biased toward either the upper or lower side of the ribbon. Such states form loops together with certain bulk states that remain in the $\PT$-symmetric phase. 
\par
Under OBC, these loops collapse into line segments,
as shown in Fig.~\ref{fig:appenA}(b). The summed spatial distributions of the green (red) dots are shown in Fig.~\ref{fig:appenA}(e) [Fig.~\ref{fig:appenA}(f)].
The green (red) states localize on  the right (left) boundary, exhibiting  a bipolar skin effect~\cite{bipolar1,bipolar2,bipolar3}, where
distinct groups of states accumulate at opposite boundaries.
\par
The physical origin of the NHSE can be understood as follows. Consider, for example,  the green states in Fig.~\ref{fig:appenA}(b). Under OBC, these states arise from the hybridization between a $\PT$-broken-phase state that is  predominantly localized near the upper side and propagates to the right [Fig.~\ref{fig:appenA}(c)], and a $\PT$-symmetric bulk state that is distributed more uniformly and propagates to the left. 
Since the upper side carries gain, the right-propagating component is amplified. Therefore, these states are localized on the right side.
\par
These results demonstrate that when $|\ga_1|=|\ga_2|>\ga_0$, some bulk states enter the $\PT$-broken phase and the bulk NHSE emerges even along
the $\PT$-symmetric line.

\section{\label{appen:NHSEinter}detailed physical interpretation of the NHSE}
\subsection{\label{Llossinter}Lower-edge loss}
\begin{figure}[tb]
    \centering
    \includegraphics[width=0.99\linewidth]{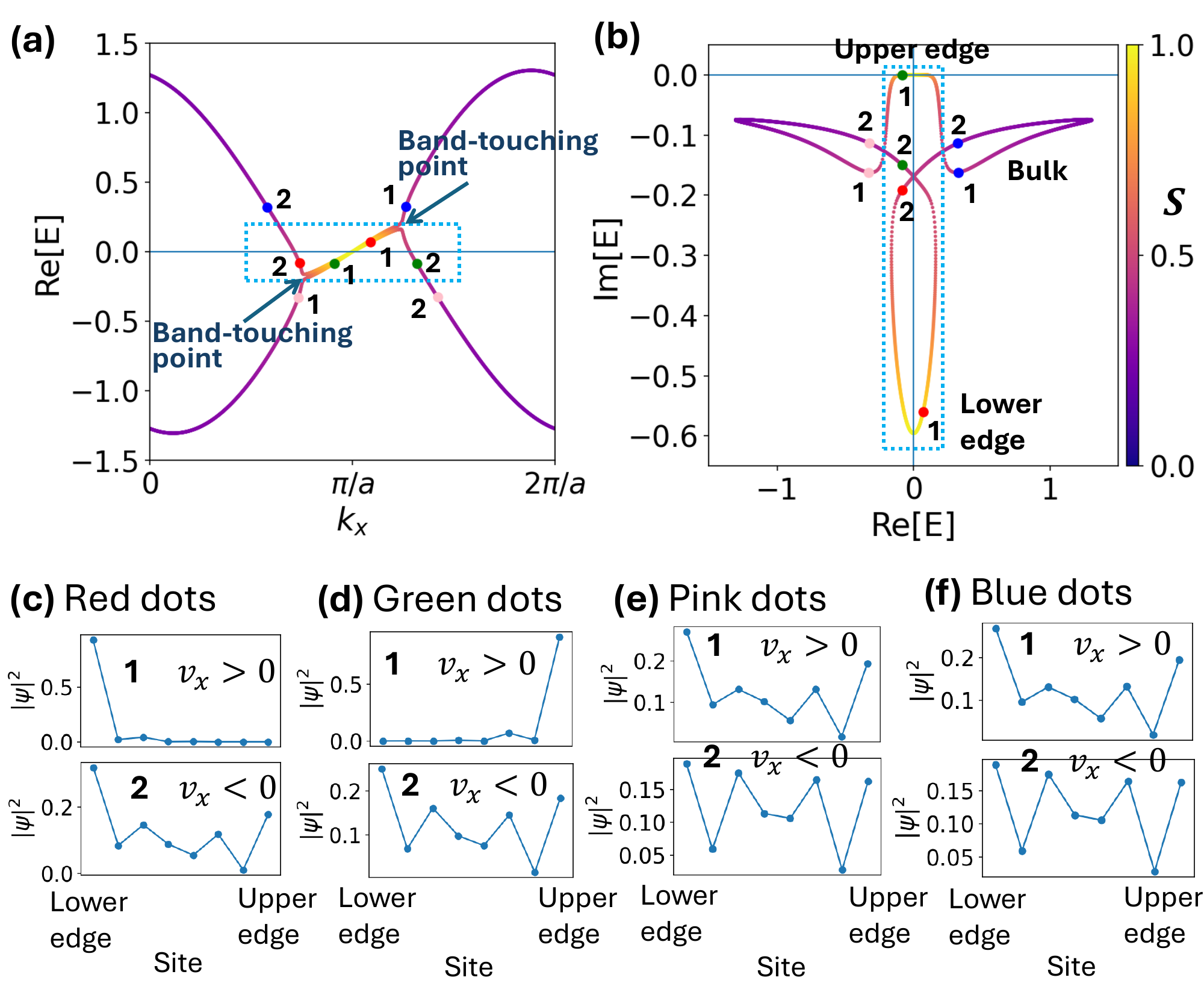}
    \caption{Spatial profiles of representative eigenstates under PBC for $H(\gamma_1=-0.6,\gamma_2=0)$. (a) Band structures of $\mathrm{Re}[E_1]$ and $\mathrm{Re}[E_2]$, corresponding to two branches highlighted in Fig.~\ref{fig:Lloss}(b). (b) Complex-energy spectrum of $E_1$ and $E_2$. The eigenenergies enclosed by the blue dashed rectangle correspond to those enclosed in (a). Brighter colors indicate stronger localization at the upper and lower edges. (c) Spatial profiles of the two red dots in (a) and (b).  (d) Spatial profiles of the two green dots in (a) and (b). (e) Spatial profiles of the two pink dots in (a) and (b). (f) Spatial profiles of the two blue dots in (a) and (b). In (c)-(f), the displayed region
    corresponds to the sites within the unit cell shown in Fig.~\ref{fig:Lloss}(a). $v_x$ denotes the group velocity along the $x$ direction of the state shown.
    }
    \label{fig:Lloss2}
\end{figure}
In this appendix, we provide a detailed physical interpretation of the results shown in Sec.~\ref{subsec:L} ($\ga_1=-0.6,\,\ga_2=0$).
\par
We first discuss the origin of the NHSE in the lower-edge states.
Figures~\ref{fig:Lloss2}(a) and~\ref{fig:Lloss2}(b) show the band diagram and the energy spectrum, respectively, for the $E_1(k_x)$ and $E_2(k_x)$ bands. These panels correspond to Figs.~\ref{fig:Lloss}(b) and~\ref{fig:Lloss}(d), respectively.
The red dots in Fig.~\ref{fig:Lloss2}(a) and \ref{fig:Lloss2}(b) belong to the lower loop in Fig.~\ref{fig:Lloss2}(b). The corresponding eigenstates within a unit cell defined in Fig.~\ref{fig:Lloss}(a) for dot 1 (lower-edge state) and dot 2 (bulk state) are shown in Fig.~\ref{fig:Lloss2}(c). Since the lower edge consists only of loss sites, $\ga_{\mathrm{eff}}^{\mathrm{L\text{-}edge}}<0$. From the bottom panel of Fig.~\ref{fig:Lloss2}(c), although the bulk state also has weight on the lower edge, its weight is smaller than that of the lower-edge state, leading to
$\ga_{\mathrm{eff}}^{\mathrm{Bulk}}>\ga_{\mathrm{eff}}^{\mathrm{L\text{-}edge}}.$
Since the lower-edge states propagate to
the right whereas the corresponding bulk states
propagate to the left, this imbalance results in
relative attenuation of the edge 
states toward the right and relative amplification of the bulk states toward the left.
Under OBC, these states hybridize and localize at the left boundary [Fig.~\ref{fig:Lloss}(f)].
Because the lower-edge states are strongly localized at the edge and the number of bulk states in this group is smaller than that of the edge states, the state density on the edge is much larger. Consequently, the bulk density in Fig.~\ref{fig:Lloss}(f) is barely visible.
\par
We now turn to the NHSE observed in the upper-edge states, which is called the nonlocal antichiral skin effect. The green dots in Fig.~\ref{fig:Lloss2}(a) and \ref{fig:Lloss2}(b) belong to the upper loop in Fig.~\ref{fig:Lloss2}(b). The corresponding eigenstates for dot 1 (upper-edge state) and dot 2 (bulk state) are shown in Fig.~\ref{fig:Lloss2}(d).
Although no gain or loss is directly applied to
the upper edge, the bulk states have finite 
weight on the lossy lower edge, giving 
$\ga_{\mathrm{eff}}^{\mathrm{Bulk}}<0$,
while $\ga_{\mathrm{eff}}^{\mathrm{U\text{-}edge}}\approx 0$.  Thus, $\ga_{\mathrm{eff}}^{\mathrm{Bulk}}<\ga_{\mathrm{eff}}^{\mathrm{U\text{-}edge}}$. Unlike the lower-edge case, here the upper-edge states are relatively amplified toward the right, whereas the bulk states are relatively attenuated toward the left. Therefore, under OBC the NHSE occurs, and the upper-edge states hybridize with some bulk states that are localized on the right side, as shown in Fig.~\ref{fig:Lloss}(g). 
\par
Finally, we explain the NHSE occurring in the bulk states other than those enclosed by the blue dashed rectangle in Fig.~\ref{fig:Lloss2}(a), which are shown in Fig.~\ref{fig:Lloss}(h). Under PBC, these states include bulk states propagating in both the left and right directions. However, for the states with the same $\mathrm{Re}[E]$, the states propagating to the right [e.g., the top panels of Figs.~\ref{fig:Lloss2}(e) and \ref{fig:Lloss2}(f)] have larger weight on the lower edge than those propagating to the left [e.g., the bottom panels of Figs.~\ref{fig:Lloss2}(e) and \ref{fig:Lloss2}(f)].  For eigenstates with the same $\mathrm{Re}[E]$, we define the effective gain/loss [Eq.~\eqref{eff}] of right- and left-propagating states as $\ga_{\mathrm{eff}}^{\mathrm{right}}$ and $\ga_{\mathrm{eff}}^{\mathrm{left}}$, respectively. Since there are loss sites at the lower edge, we obtain $\ga_{\mathrm{eff}}^{\mathrm{right}}<\ga_{\mathrm{eff}}^{\mathrm{left}}$. The right-attenuating state and the left-amplifying state are coupled under OBC, and the eigenstates become localized on the left side. This is the physical explanation of the NHSE induced solely by bulk states [Fig.~\ref{fig:Lloss}(h)]. As in the emergence of antichiral skin modes, this effect is also induced by the presence of $\phi$. 
In Fig.~\ref{fig:anti}(c), the two Dirac points are shifted in the vertical direction by $\phi$ \cite{MHM}. For finite $N_y$, the band-touching points of $E_1$ and $E_2$ (corresponding to the Dirac points) are also shifted by $\phi$ [Fig.~\ref{fig:Lloss2}(a)]. As a result, for states with the same $\mathrm{Re}[E]$, eigenstates with $v_x>0$ are located closer to the band-touching points. In addition, in the $E_1$ and $E_2$ bands, bulk states near the band-touching points tend to have larger weight at the edges. Therefore, eigenstates with $v_x>0$ have larger weight on the edges.
\par

\subsection{\label{Ugaininter}Upper-edge gain}
We now provide a physical interpretation of the results shown in Sec.~\ref{subsec:U} ($\ga_1=0,\ga_2=+0.6$). The underlying mechanism is essentially the same as in the case of $H(\ga_1=-0.6,\ga_2=0)$. 
\par
First, we consider the NHSE of the upper-edge states and the lower-edge states. Since gain is applied only to the upper edge, the effective gain/loss satisfies
$\ga_{\mathrm{eff}}^{\mathrm{U\text{-}edge}}>\ga_{\mathrm{eff}}^{\mathrm{Bulk}}>\ga_{\mathrm{eff}}^{\mathrm{L\text{-}edge}}$. This hierarchy is analogous to that in the previous subsection.
Consequently, the localization directions of the edge states 
are the same as those in the lower-edge-loss case.
\par
\begin{figure}[tb]
    \centering
    \includegraphics[width=0.99\linewidth]{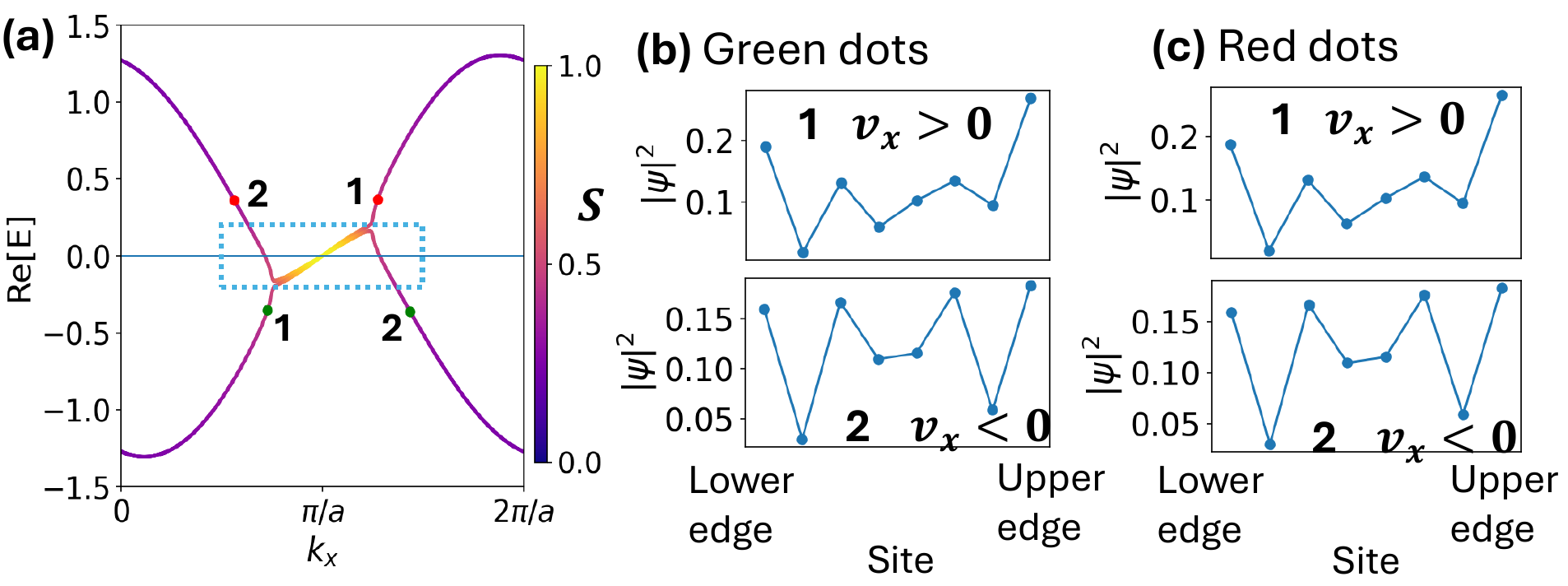}
    \caption{Spatial profiles of representative eigenstates under PBC for $H(\gamma_1=0,\gamma_2=+0.6)$. (a) Band structures of $\mathrm{Re}[E_1]$ and $\mathrm{Re}[E_2]$. (b) Spatial profiles of the two green dots in (a). (c) Spatial profiles of the two red dots in (a).
    In (b) and (c), the displayed region corresponds to the sites within the unit cell shown in Fig.~\ref{fig:Ugain}(a).
    $v_x$ denotes the group velocity along the $x$ direction of the state shown.
    }
    \label{fig:Ugain2}
\end{figure}
We next examine the NHSE of the bulk states. As 
shown in Fig.~\ref{fig:Ugain2},
 eigenstates with $v_x>0$ have larger weights near the edges than those with $v_x<0$. 
 Because the upper edge now carries gain, right-propagating states experience stronger amplification
 than left-propagating states, leading to
  $\ga_{\mathrm{eff}}^{\mathrm{right}}>\ga_{\mathrm{eff}}^{\mathrm{left}}$. Thus, the bulk states are localized on the right side of the system.

\section{\label{appenB}Bulk NHSE in bands other than $E_1$ and $E_2$}
In this Appendix, we present bulk eigenstates of $H(\ga_1=-0.6,\ga_2=0)$ that belong to bands other than $E_1$ and $E_2$ shown in Fig.~\ref{fig:Lloss}(b). These bulk states also exhibit the NHSE due to the finite ribbon width and the presence of edge loss.
\begin{figure}[!htb]
    \centering
    \includegraphics[width=0.99\linewidth]{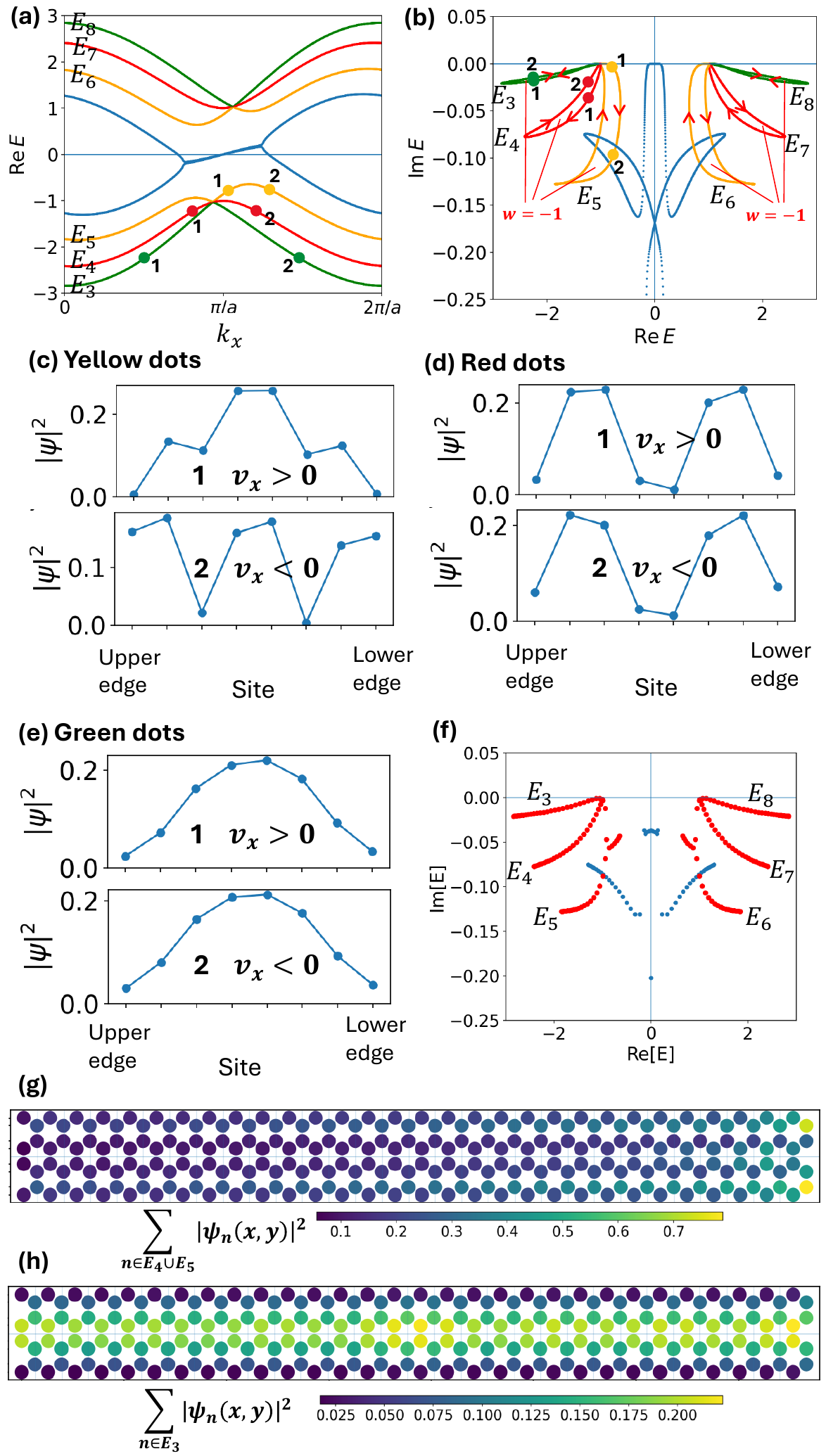}
    \caption{Bulk NHSE in bands other than $E_1$ and $E_2$ for $H(\ga_1=-0.6,\ga_2=0)$. 
    (a) Band structure of $\mathrm{Re}[E]$ under PBC along the $x$ direction, highlighting the bands $E_3-E_8$. 
    (b) Complex-energy spectrum under PBC. 
    The red arrows indicate the direction of each loop as $k_x$ increases, and $w$ denotes the point-gap winding number. 
    (c) Spatial profiles of the two yellow dots in (a) and (b). 
    (d) Spatial profiles of the two red dots in (a) and (b). 
    (e) Spatial profiles of the two green dots in (a) and (b).
    In (c)-(e), $v_x$ denotes the group velocity along the $x$ direction of the state shown.
    (f) Energy spectrum under OBC along $x$. 
    Red dots indicate eigenenergies originating from $E_3-E_8$. 
    (g) Spatial distribution of the summed bulk states originating from $E_4$ and $E_5$. 
    (h) Spatial distribution of the summed bulk states originating from $E_3$. 
    In (g) and (h), brighter colors indicate higher state density. 
    In (f)-(h), the system size is $N_x=30$.
    }
    \label{fig:B}
\end{figure}
\par
We label the relevant bands as $E_3$--$E_8$ as shown in Fig.~\ref{fig:B}(a). The corresponding complex-energy
spectrum under PBC is shown in
Fig.~\ref{fig:B}(b). Bands $E_4$, $E_5$, $E_6$, and $E_7$ form point gaps with $w=-1$, implying localization on the right boundary under OBC. 
As an example, Fig.~\ref{fig:B}(g) shows the summed
spatial distribution of the OBC states originating from $E_4$ and $E_5$, confirming their localization on the right side.
\par
On the other hand, as seen in Fig.~\ref{fig:B}(b), $E_3$ and $E_8$ exhibit only very small point gaps. In fact, Fig.~\ref{fig:B}(h), which shows the summed OBC states  of $E_3$, indicates that these states remain nearly extended. Although a  weak NHSE is expected in principle, it is practically negligible.
\par
The physical origin of this behavior can be understood from
the spatial structure of the eigenstates under PBC. For eigenstates with the same $\mathrm{Re}[E]$, we define the effective gain/loss [Eq.~\eqref{eff}] of right- and left-propagating states as $\ga_{\mathrm{eff}}^{\mathrm{right}}$ and $\ga_{\mathrm{eff}}^{\mathrm{left}}$, respectively. As shown in Figs.~\ref{fig:B}(c) and \ref{fig:B}(d), 
for states with the same $\mathrm{Re}[E]$ in
 $E_4$ and $E_5$,  those with $v_x<0$ have  larger weight near the lower edge than those with $v_x>0$. Since the lower edge carries loss, this leads to $\ga_{\mathrm{eff}}^{\mathrm{left}}<\ga_{\mathrm{eff}}^{\mathrm{right}}$. 
 Under OBC, this imbalance causes localization on the right boundary.
 \par
 By contrast, as shown in Fig.~\ref{fig:B}(e), the states in $E_3$ are distributed predominantly around the center of the ribbon and have negligible weight on the edges. Consequently, 
the edge loss has little influence on these states,
and the NHSE is suppressed.
\par
Combining these results with those in Sec.~\ref{subsec:L}, we conclude that the bulk states originating from $E_1$ and $E_2$ localize on the left boundary, whereas the other bulk states either localize on the right boundary or remain nearly extended. 
This coexistence of oppositely localized bulk modes constitutes a bipolar skin effect. 
\par
$H(\ga_1=-0.6,\ga_2=0)$ belongs to phase II, which is defined in Sec.~\ref{subsec:phase}.
 In phase I, 
 the localization direction of each band is reversed 
 under the $\PT$ transformation.

\section{\label{appenlargeNy}Lower-edge loss for large $N_y$}
\begin{figure}
    \centering
    \includegraphics[width=0.99\linewidth]{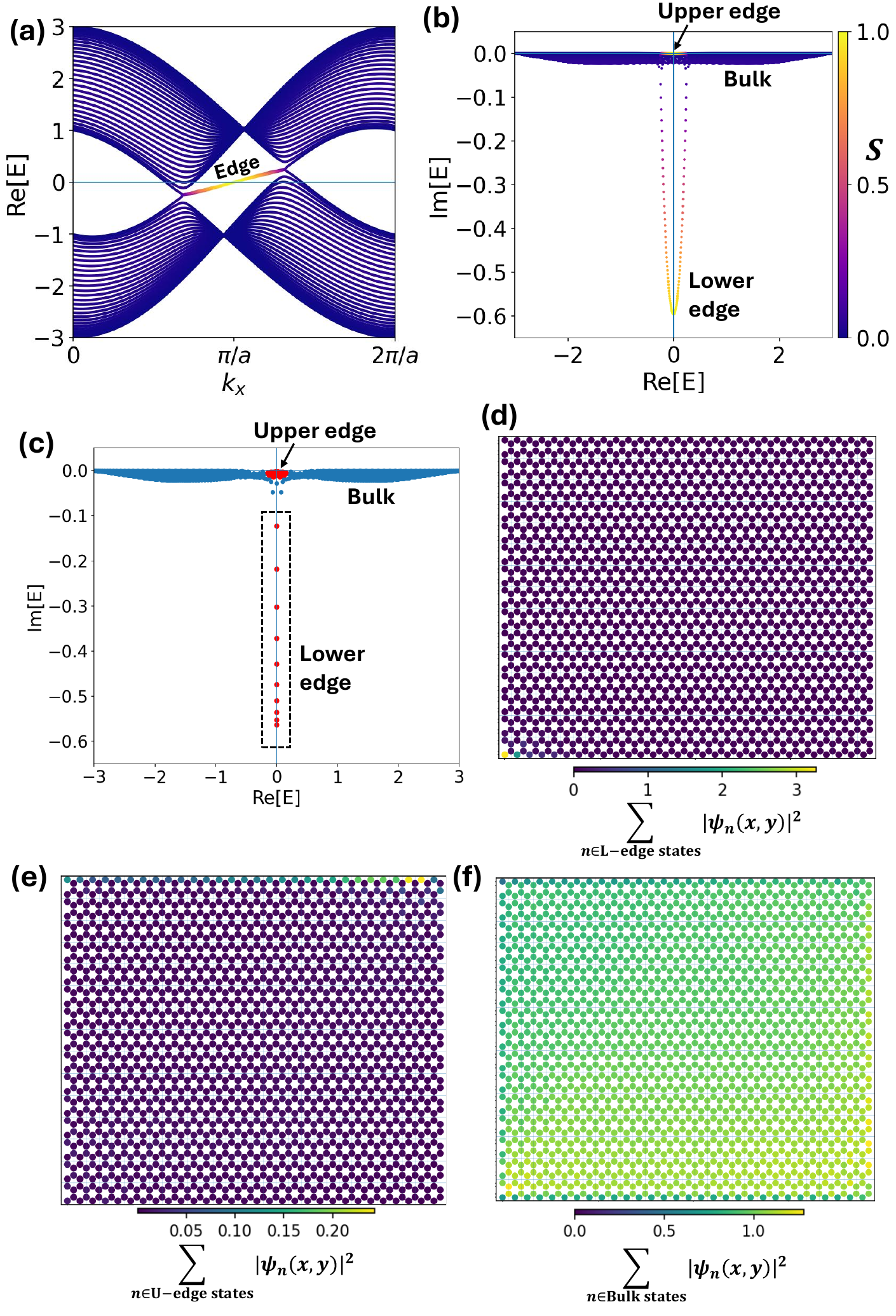}
    \caption{
    Results for $H(\ga_1=-0.6,\ga_2=0)$ with large $N_y$.
    (a) Band structure of $\mathrm{Re}[E]$ under PBC along the $x$ direction. 
    (b) Complex-energy spectrum under PBC.
    Brighter colors indicate stronger localization at the upper and lower edges.
    In (a) and (b), the system size is $N_y=60$.
    (c) Energy spectrum under OBC along $x$. 
    (d) Spatial distribution of the summed lower-edge states selected from (c). 
    (e) Spatial distribution of the summed upper-edge states selected from (c). 
    (f) Spatial distribution of the summed bulk states selected from (c).
    In (c)-(f), the system size is $N_x=30, \, N_y=60$.
    }
    \label{fig:wide}
\end{figure}
To examine how the NHSE discussed in this paper 
depends on the ribbon width, 
we present numerical results for a wider modified
Haldane nanoribbon. Specifically, we consider
$H(\ga_1=-0.6,\ga_2=0)$, corresponding to loss
applied only to the lower edge, with $N_y=60$.
\par
Figures~\ref{fig:wide}(a) and \ref{fig:wide}(b) show the band structure and the energy spectrum under PBC
along the $x$ direction, respectively.  As shown in Fig.~\ref{fig:wide}(b),  all bulk states and the upper-edge states are distributed close to the real axis and occupy only a small region in the complex-energy plane, whereas the lower-edge states still form a pronounced loop. This behavior reflects the fact that the lower-edge states are directly affected by the loss introduced at the lower edge, whereas the upper-edge and bulk states are only weakly affected by it and therefore remain nearly Hermitian. Since the point gap opened by the bulk states is extremely small, the resulting NHSE is expected to be negligible.
\par
Figure~\ref{fig:wide}(c) shows the energy spectrum under OBC, while Figs.~\ref{fig:wide}(d), \ref{fig:wide}(e), and \ref{fig:wide}(f) show the summed probability distributions of the lower-edge, upper-edge, and bulk states, respectively. As shown in Fig.~\ref{fig:wide}(e), the upper-edge states extend over the entire upper edge, while being biased toward the right side. This indicates that the nonlocal antichiral skin effect still exists even in a relatively wide ribbon. However, it is much weaker than that in Fig.~\ref{fig:Lloss}(g). As shown in Fig.~\ref{fig:wide}(f), the bulk states are extended over the entire system, with almost no observable NHSE.

\nocite{*}

\bibliography{main}% Produces the bibliography via BibTeX.

\end{document}